%% file: Apr02_2020_FPL_2020_Submitted_Version_arXiv.tex
\documentclass[conference,draftcls,onecolumn,letter]{IEEEtran}
\usepackage{xcolor}

\include{import_packages_and_definitions}
\usepackage{stfloats}
\usepackage{balance}

\def\BibTeX{{\rm B\kern-.05em{\sc i\kern-.025em b}\kern-.08em
    T\kern-.1667em\lower.7ex\hbox{E}\kern-.125emX}}

\begin{document}
\title{Hardware Trojan with Frequency Modulation}

\author{\IEEEauthorblockN{Ash LUFT, Mihai SIMA, and Michael McGUIRE}
\IEEEauthorblockA{Department of Electrical and Computer Engineering,
University of Victoria, British Columbia, Canada\\
Email: aluft@uvic.ca, msima@ece.uvic.ca, mmcguire@uvic.ca}}



\IEEEpubid{\makebox[\columnwidth]{978-1-0000-0000-0/20/\$31.00~\copyright~2020~IEEE\hfill}
\hspace{\columnsep}\makebox[\columnwidth]{}}

\maketitle
\IEEEpubidadjcol

\section{Introduction}


The globalization of the economy is offering the choice of importing designs and
third\nobreakdash-party Intellectual Property (IP) cores from multiple vendors located
all over the world.  An IP can be compromised in an untrusted design or fabrication facility
through the insertion of malicious circuitry referred to as a Hardware
Trojan\cite{Banga_Jul10,Zhang_Jun11}, which triggers
a malfunction under rare circuit conditions\cite{Wolff_Mar08}.  Due to the stealthy
nature of hardware Trojans, their detection is challenging\cite{Chakraborty_Nov09}.

Three methods for the detection of hardware Trojans have been proposed:
(i)~state\nobreakdash-change detection through Malicious 
Circuit Activation (MCA),
which aims to determine under which conditions the malicious hardware 
is activated\cite{Tehranipoor_JanFeb10}, or alternatively Unused\nobreakdash-Circuit 
Identification (UCI)\cite{Hicks_May10}, which aims to determine which
circuitry is not activated under normal operating conditions;
(ii)~Side\nobreakdash-Channel Analysis (SCA), in which circuit parameters (such as power 
consumption or propagation delay) are estimated or measured to determine the Trojan's
contribution\cite{Lin_Sep09,Rad_Dec10,Zhang_Mar11}; 
and (iii)~use of Monitoring Architectures (MA), which will indicate if the original 
layout of the integrated circuit design has been changed\cite{Zhang_Mar11,Rilling_Oct11}.

To facilitate research in this area and improve the detection of malicious alterations, a
better understanding of what hardware Trojans would look like and what impact they would
incur to a circuit or an IP is needed\cite{Jin_Jul09,Becker_Oct11}.  In order to show 
the weaknesses of current
Trojan detection methods, a logic family with frequency modulation for building hardware
Trojans that can evade the first two major detection methods is described.  Specifically, 
since the Trojan trigger circuitry's state will never stay constant during 'normal' operation, 
UCI\nobreakdash-class analyses are evaded.  The power consumption of the Trojan's trigger 
circuitry can be balanced into a constant value (thus concealed) with minimal design
effort and supplementary hardware resources. This ensures that the Trojan can evade
side\nobreakdash-channel analysis.
The hardware Trojan logic family is designed to infect synthesizable IPs for
FPGAs, for which an original physical placement does not exist.  This makes the third
major detection method, namely the use of monitoring architectures, immaterial for the
rest of the presentation.  

The proposed hardware Trojan has a lightweight implementation, as it does not require
more Look\nobreakdash-Up Tables (LUT) than standard logic, but only extra
Flip\nobreakdash-Flops (FF) which are abundantly available in FPGAs.  Electrically it 
can be classified as Always\nobreakdash-On\cite{Tehranipoor_JanFeb10,Karri_Oct10}, so 
that it can evade the UCI detection\cite{Hicks_May10}.  Functionally it is a
condition\nobreakdash-based Trojan\cite{Tehranipoor_JanFeb10}, which means its payload is
activated under a very specific condition, namely a specific sequence of events. 
Such a Trojan is therefore unlikely to be activated during functional
tests\cite{Tehranipoor_JanFeb10} so it will evade the MCA detection.  The payload of the
proposed Trojan does not directly control output pins since that would make it easier to
detect the Trojan.  Instead, the payload communicates in a controlled fashion through 
a power consumption side\nobreakdash-channel. The proposed logic family is
synthesizable; thus, it can be ported to Application\nobreakdash-Specific Integrated 
Circuits (ASIC) with minimal effort. 

To summarize, a logic family which does not exhibit idle states 
(thus, it is always active and evades the first detection method), makes the power 
consumption independent relative to the processed data and/or operations (thus, it 
conceals the power consumption and evades the second detection method), and can 
readily be implemented with FPGA primitives is proposed.
The contributions of this paper are as follows.
\begin{itemize}
\item Logic family with Frequency Modulation (FM), which does not exhibit idle
states, thus being able to escape detection schemes based on unused\nobreakdash-circuit
identification.
\item Trojan FM~logic circuit augmentation to evade detection schemes based on power
consumption.
\item Payload which communicates in a controlled fashion through
a power consumption side\nobreakdash-channel.
\item Detection approaches for hardware Trojans built in the proposed logic family 
with frequency modulation.
\end{itemize}

\section{Background}
\label{sect:background}

\subsection{FPGA Architecture}
\label{subsect:FPGA_architecture}

Modern FPGA architectures consist of five types of modules: 
Configurable Logic Blocks (CLB), Digital Signal Processing (DSP) units, Block
Random Access Memories (BRAM), I/O~Blocks (IOB), and a configurable Interconnection
Network.  The CLBs are part of the fine\nobreakdash-grained fabric and are organized
as a two dimensional array, where each CLB includes a configurable Look-Up Table (LUT) to
implement bit\nobreakdash-level logic functions, carry logic to support
arithmetic operations such as binary\,/\,ternary adders\cite{Simkins_Sep07}, dedicated 
multiplexors, and Flip\nobreakdash-Flops (FF).  Software support includes
macros and primitives, such as \textsl{CARRY4} which concatenates four LUTs to build a 
4\nobreakdash-bit binary\,/\,ternary adder, \textsl{MUXF7} and 
\textsl{MUXF8} which instantiate 2\nobreakdash-to\nobreakdash-1 multiplexors, and
\textsl{BRLSHFT4} and \textsl{BRLSHFT8} which instantiate 4\nobreakdash-bit and 
8\nobreakdash-bit barrel shifters, etc.  The DSP units are part of the 
coarse\nobreakdash-grained fabric and implement large two's\nobreakdash-complement 
multipliers and acumulators.  The configurable interconnection network
connects these modules together to implement digital circuits.

Modern FPGAs also integrate Block Random Access Memories (BRAM) on chip.  For example, 
the BRAM in the Virtex\nobreakdash-7 family\cite{Xilinx_Virtex7_Feb19_a} can operate as either
one $36\, \mathrm{Kb}$ dual\nobreakdash-port RAM
or two independent $18\, \mathrm{Kb}$ dual\nobreakdash-port RAMs.
These BRAMs can be used as large storage areas or large LUTs with
multiple outputs to implement logic functions.

Modern FPGAs offer a large number of flip\nobreakdash-flops of different 
types along with their software primitives\cite{Xilinx_Virtex7_Dec18_a}. \textsl{FDSE} and 
\textsl{FDRE} are D~flip\nobreakdash-flops with clock enable (CE) and synchronous set 
(S) and reset (R), respectively. When the (S)\,/\,(R) input is High, the (Q) output
is set\,/\,reset on the Low\nobreakdash-to\nobreakdash-High clock (C) transition. 
When (S)\,/\,(R) is Low and (CE) is high the data on the (D) input is loaded into the
flip\nobreakdash-flop on the Low\nobreakdash-to\nobreakdash-High clock (C) transition.

The \textsl{LUT6} primitive refers to a 6\nobreakdash-input, 1\nobreakdash-output 
look\nobreakdash-up table (LUT) that can either act as an asynchronous 64\nobreakdash-bit 
ROM (that is, with a 6\nobreakdash-bit address bus) or implement an arbitrary 
6\nobreakdash-input logic function. LUTs and flip\nobreakdash-flops are the most basic 
logic building blocks in FPGA circuitry.  They will be used in implementing the
Trojan logic with frequency modulation.

\section{Hardware Trojans}
\label{sect:hardware_trojans}

A hardware Trojan is characterized by its activation mechanism, which is referred to as a
\textbf{trigger}, and the malicious function that it implements, which is referred to as 
a \textbf{payload}\cite{Wang_Jun08,Tehranipoor_JanFeb10}. The activation of a
Trojan is a statistically rare event, such as the end users or the standard verification
tests during manufacturing will very likely not trigger it. In this respect, a hardware
Trojan is a stealthy circuit\cite{Wolff_Mar08}.

Based on their trigger mechanism, malicious circuits are classified as follows.
\textbf{Condition\nobreakdash-based} Trojans are inactive until a specific condition is
met (for example, the attacker provides a special input or a particular value occurs on the
data bus). \textbf{Always\nobreakdash-on} Trojans operate continuously, but they are 
inserted on nodes which are rarely exercised\cite{Becker_Oct11}.

The payload can take many forms. In one scenario, sensitive data (such as an encryption key) 
can be sent to the attacker through an output port, which normally transmits plain text.
In another scenario it is possible to send the sensitive data to the attacker through side 
channels in which power consumption or some electromagnetic radiation originating from 
the chip can be modulated with the encryption key. The payload can also compromise the
operation of the circuit, or it can even physically destroy the chip.

As mentioned, there are three major detection methods.  The first
method, the \textbf{Malicious Circuit Activation} (MCA), is based on applying stimuli 
at the circuit's input to determine under which conditions the malicious hardware is 
activated\cite{Tehranipoor_JanFeb10}.  Since hardware Trojans are stealthy circuits,
this method can be very time consuming; thus, it is ineffective at consistently
detecting trigger states.  The \textbf{Unused 
Circuit Identification} (UCI) method proposed by Hicks~\etal ~is complementary to 
MCA\cite{Hicks_May10}. It flags as 
suspicious those circuits that are not activated under normal operating conditions or 
by design verification tests\cite{Hicks_May10}. Waksman, Souzzo, and 
Sethumadhavan proposed a UCI\nobreakdash-class 
analysis for
nearly\nobreakdash-unused circuit identification\cite{Waksman_Nov13}, which aims
to find gates and their inputs that very rarely impact the outputs of a logic circuit.
In the same class of unused circuit identification methods, VeriTrust, which was proposed
by Zhang~\etal, performs a logic analysis to identify the unused inputs of a logic 
circuit\cite{Zhang_Jul15}. Malicious hardware able 
to evade the UCI analysis are discussed next.

Sturton~\etal~proposed to create a circuit in which no pair of dependent signals 
are always equal under the non\nobreakdash-trigger condition\cite{Sturton_May11}. This is
equivalent to saying that the logic circuitry inscribed by these signals is not idle. In their
work the authors do not mention whether it is always possible to generate such a circuit
and with what design effort. Circuits with frequency modulation are
always active; therefore, there are not any circuit existance concerns in the proposed logic.

Zhang and Xu\cite{Zhang_Aug13} and later Zhang, Yuan, and Xu\cite{Zhang_Nov14}
have proposed implementations to evade Trojan detection through tests in the UCI class.
Rather than implementing a very rare global trigger condition, which is not activated during
standard verification but can be captured by UCI and deemed suspicious, the authors propose 
to implement sub\nobreakdash-trigger conditions that can be each covered by standard
verification.  This way, the sub\nobreakdash-trigger conditions will not be labeled as
suspicious.  Then, these sub\nobreakdash-trigger conditions are integrated using logic
operators.
However, the global trigger condition output is idle until the condition is activated.  
This is one weakness of
this technique.  A second weakness, which is acknowledged by the authors, is that it is
nearly impossible to sensitize all non\nobreakdash-trigger conditions, which in turn could
cause the Trojans to be detected by UCI techniques. To evade the UCI techniques without
exercising all non\nobreakdash-trigger conditions, the trigger condition needs to be 
carefully partitioned such that the probability of each sub\nobreakdash-trigger condition
is increased. This requirement is in contradiction with the need of 
as\nobreakdash-controllable\nobreakdash-as\nobreakdash-possible Trojan's payload,
making the design and implementation of a hardware
Trojan very challenging\cite{Zhang_Aug13}. The proposed logic family with frequency
modulation can escape the UCI tests while ensuring the flexibility of controlling 
the payload.

Krieg, Wolf, and Jantsch have proposed an RTL\nobreakdash-level Trojan\cite{Krieg_Nov16}.
Their malicious circuit is triggered under the control of the software tool at the time 
when the FPGA bitstream is generated; therefore, any test performed at the RTL level
will not raise suspicion.  The described logic family with frequency modulation 
allows the implementation of hardware Trojans at the RTL level but without the help 
of any software tool, which will allow the distribution of the IP in RTL form.

A second major detection method of malicious circuits is based on 
\textbf{side\nobreakdash-channel analysis}, in which circuit parameters, such as 
propagation delay or power consumption, are estimated or measured to determine the Trojan
contribution\cite{Lin_Sep09,Zhang_Mar11}. Hiding (or concealing) countermeasures
maintain a constant power consumption; thus, the hardware Trojan is stealthy. 
This is a problem related to attack and defense of cryptosystems based on power 
consumption\cite{Kocher_Dec99,Brier_Aug04,Mangard_10}.

In the third major detection method of malicious circuits \textbf{monitoring 
architectures} can be deployed on chip to indicate whether the original 
physical placement or layout of the integrated circuit design has been changed. Examples
in this class include the use of ring oscillators\cite{Zhang_Mar11,Rilling_Oct11}.
Recall that the hardware Trojan logic family is designed to infect synthesizable IPs for
FPGAs, for which an original physical placement not yet exists.  This makes the
monitoring architectures immaterial for this paper.

\begin{figure*}[t]
\centering
\includegraphics[scale=0.512]{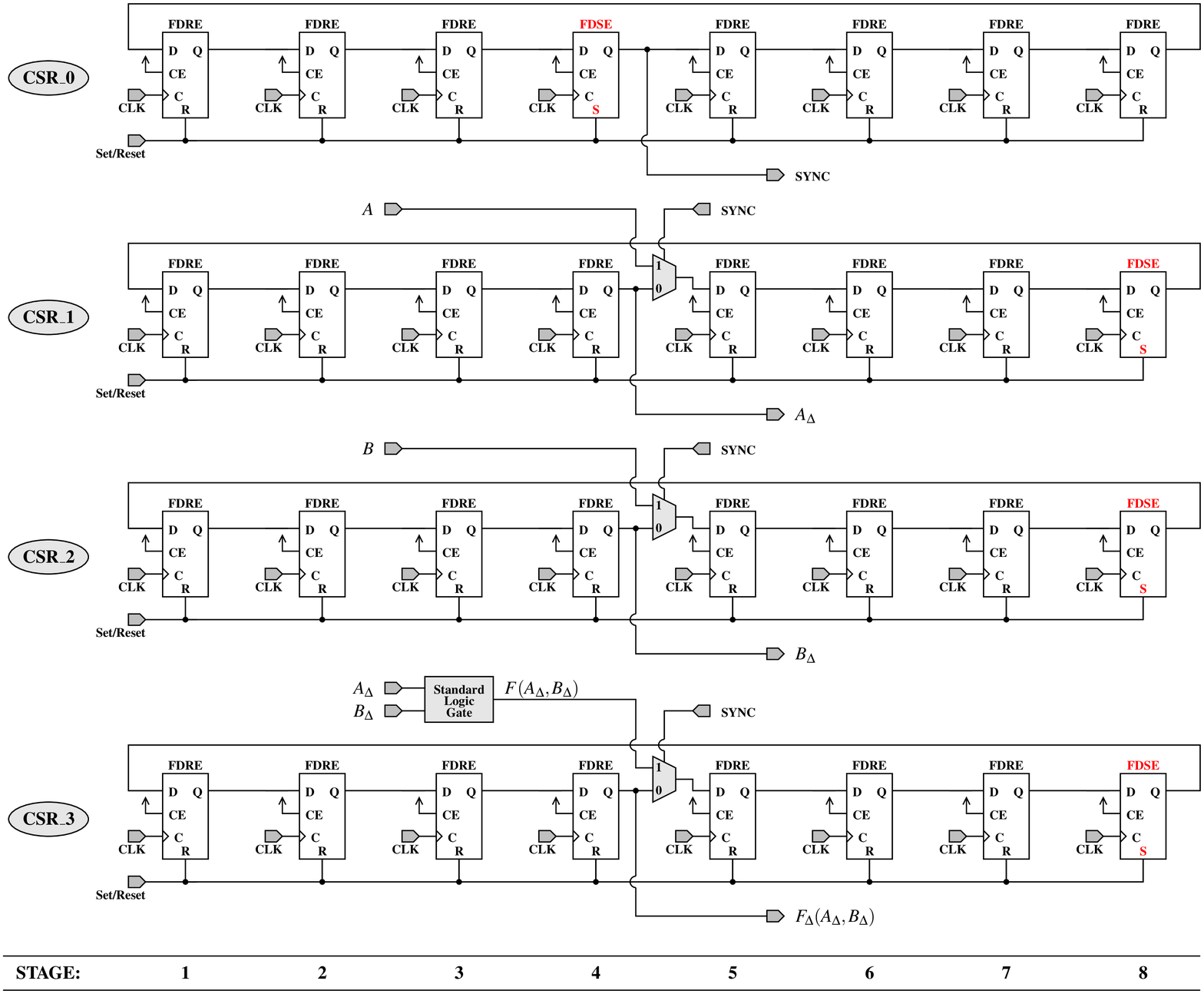}
\caption{Logic gate with frequency modulation implemented with circular shift registers.}
\label{fig:FM_logic_gate_01}
\end{figure*}

\section{Logic Family with Frequency Modulation}
\label{sect:logic_family_FM}

Similar to the Illinois Malicious Processor, which includes a state machine that 
looks for a special sequence of bytes on the data bus to activate the Trojan's 
payload\cite{King_Apr08}, the proposed hardware Trojan is activated by a specific
(long) string of processor operation codes.  Since each operation code is part of the
processor architecture, it will be sensitized during functional testing and, 
therefore, will escape a UCI\nobreakdash-class analysis. It will be understood that 
a Trojan's activation based on a string of operation codes is given in this paper 
by way of example and not by limitation. Other (long) strings of events can be conceived
to trigger the proposed Trojan.

In order to evade a UCI\nobreakdash-class analysis the output of any gate implementing the
trigger circuit must not be idle.  Synthesizable Trojans mapped onto FPGAs would
require that gates and circuits with continuous activity be implemented with standard 
primitives, which are abundant in the reconfigurable fabric. This is achieved by a
logic family with frequency modulation described below.

In standard logic, the information is encoded into the amplitude of the signal, where
Logic~\textttbf{'1'} is encoded as a large amplitude signal, $V_{\text{DD}}$, and
Logic~\textttbf{'0'} is encoded as a small amplitude signal, $\text{GND}$. In the 
proposed logic the information is encoded into the frequency of a periodic signal, where
Logic~\textttbf{'1'} is encoded as a high frequency signal, $f_1$, and 
Logic~\textttbf{'0'} is encoded as a low frequency signal, $f_0$. The signal modulation 
in frequency is implemented with a set of Circular Shift Registers
(CSR), as shown in Figure~\ref{fig:FM_logic_gate_01}. The remainder of the discussion
will consider CSRs with eight stages if not stated otherwise.
Figure~\ref{fig:logic_encoding_FM_logic_gate_01} summarizes the logic states in the
proposed logic family with frequency modulation in which $f_1 = 1/4\cdot f_{\text{CLK}}$ and 
$f_0 = 1/8\cdot f_{\text{CLK}}$.

\begin{figure}[H]
\centering
\begin{tabular}{|c|c|c|c|c|c|c|c|c|c|}
\hline
Logic    & \multicolumn{8}{c|}{Logic States in each Stage} & \multirow{2}{*}{Frequency} \\
\cline{2-9}
Encoding       & 1 & 2 & 3 & 4 & 5 & 6 & 7 & 8 & \\ 
\hline
\hline
\cellcolor{simatablegray}\color{blue}{\textttbf{0}}
  & \textttbf{0} & \textttbf{0} & \textttbf{0} & \color{blue}{\textttbf{0}}
  & \textttbf{0} & \textttbf{0} & \textttbf{0} & \cellcolor{simatablegray}\color{red}{\textttbf{1}} 
  & $1/8 \cdot f_{\text{CLK}}$ \\
\hline
\cellcolor{simatablegray}\color{blue}{\textttbf{1}} 
  & \textttbf{0} & \textttbf{0} & \textttbf{0} & \cellcolor{simatablegray}\color{blue}{\textttbf{1}} 
  & \textttbf{0} & \textttbf{0} & \textttbf{0} & \cellcolor{simatablegray}\color{red}{\textttbf{1}} 
  & $1/4 \cdot f_{\text{CLK}}$ \\
\hline
\end{tabular}
\caption{Frequency modulation logic stage encoding.}
\label{fig:logic_encoding_FM_logic_gate_01}
\end{figure}

The implementation of a 2\nobreakdash-input logic gate with Frequency Modulation (FM) is
shown in Figure~\ref{fig:FM_logic_gate_01}. The circular shift register CSR\_0 generates
the synchronization signal, $\boldsymbol{SYNC}$. This is achieved by a FDSE
flip\nobreakdash-flop on Stage~4. The other three circular shift registers, CSR\_1,
CSR\_2, and CSR\_3, generate the intermediate and final signals. They each
have a FDSE flip\nobreakdash-flop on Stage~8. The initial state that occurs after the RESET
signal is activated is shown in Figure~\ref{fig:reset_states_FM_logic_gate_01}.

\begin{figure}[H]
\centering
\begin{tabular}{|c|c|c|c|c|c|c|c|c|c|}
\hline
Shift    & \multicolumn{8}{c|}{Logic States in each Stage} & \multirow{2}{*}{Frequency} \\
\cline{2-9}
Register       & 1 & 2 & 3 & 4 & 5 & 6 & 7 & 8 & \\ 
\hline
\hline
CSR\_0 & \textttbf{0} & \textttbf{0} & \textttbf{0} & \color{red}{\textttbf{1}} 
       & \textttbf{0} & \textttbf{0} & \textttbf{0} & \textttbf{0} 
       & not relevant \\
\hline
CSR\_1 & \textttbf{0} & \textttbf{0} & \textttbf{0} & \textttbf{0} 
       & \textttbf{0} & \textttbf{0} & \textttbf{0} & \color{red}{\textttbf{1}} 
       & $1/8 \cdot f_{\text{CLK}}$ \\
\hline
CSR\_2 & \textttbf{0} & \textttbf{0} & \textttbf{0} & \textttbf{0} 
       & \textttbf{0} & \textttbf{0} & \textttbf{0} & \color{red}{\textttbf{1}} 
       & $1/8 \cdot f_{\text{CLK}}$ \\
\hline
CSR\_3 & \textttbf{0} & \textttbf{0} & \textttbf{0} & \textttbf{0} 
       & \textttbf{0} & \textttbf{0} & \textttbf{0} & \color{red}{\textttbf{1}} 
       & $1/8 \cdot f_{\text{CLK}}$ \\
\hline
\end{tabular}
\caption{Reset states for shift registers shown in Figure~\ref{fig:FM_logic_gate_01}.}
\label{fig:reset_states_FM_logic_gate_01}
\end{figure}

The input signals $\boldsymbol{A}$ and
$\boldsymbol{B}$ are first converted from standard logic to FM~logic. After eight clock
cycles, the delayed signals $\boldsymbol{A}_{\boldsymbol{\Delta}}$ and 
$\boldsymbol{B}_{\boldsymbol{\Delta}}$ become available to drive the standard logic gate 
(e.g., AND, NAND, OR, NOR, XOR, XNOR, etc.) to generate 
$\boldsymbol{F}(\boldsymbol{A}_{\boldsymbol{\Delta}},\boldsymbol{B}_{\boldsymbol{\Delta}})$.
After another eight clock cycles the output $\boldsymbol{F}_{\boldsymbol{\Delta}}
(\boldsymbol{A}_{\boldsymbol{\Delta}},\boldsymbol{B}_{\boldsymbol{\Delta}})$ is generated.
Figure~\ref{fig:FM_OR_gate_01} summarizes the operation of the FM~OR gate. The standard
logic values are combined in Stage~4, and the result will be stored in Stage~5 in the
output shift register. It is apparent that the latency of the FM gate is 8~stage delays,
which is needed for the result to make a complete turn in the circular shift register.

\begin{figure}[H]
\centering
\begin{tabular}{|c|c|c|c|c|c|c|c|c|c|}
\hline
\multirow{2}{*}{Signals}
 & \multicolumn{8}{c|}{Logic States in each Stage} & \multirow{2}{*}{Frequency} \\
\cline{2-9}
 & 1 & 2 & 3 & 4 & 5 & 6 & 7 & 8 & \\ 
\hline
\hline
Sync
      & \textttbf{0} & \textttbf{0} & \textttbf{0} & \color{red}{\textttbf{1}} 
      & \textttbf{0} & \textttbf{0} & \textttbf{0} & \textttbf{0} 
      & not relevant \\
\hline
\hline
{\bf\boldmath $A_{\Delta} = 0$} 
      & \textttbf{0} & \textttbf{0} & \textttbf{0} & \cellcolor{simatablegray}\color{blue}{\textttbf{0}} 
      & \textttbf{0} & \textttbf{0} & \textttbf{0} & \color{red}{\textttbf{1}} 
      & $1/8 \cdot f_{\text{CLK}}$ \\
\hline
{\bf\boldmath $B_{\Delta} = 1$}
      & \textttbf{0} & \textttbf{0} & \textttbf{0} & \cellcolor{simatablegray}\color{blue}{\textttbf{1}} 
      & \textttbf{0} & \textttbf{0} & \textttbf{0} & \color{red}{\textttbf{1}} 
      & $1/4 \cdot f_{\text{CLK}}$ \\
\hline
\hline
{\bf\boldmath $F_{\Delta 1} = 1$}
      & \color{red}{\textttbf{1}} & \textttbf{0} & \textttbf{0} & \textttbf{0} 
      & \color{blue}{\textttbf{1}} & \textttbf{0} & \textttbf{0} & \textttbf{0} 
      & $1/4 \cdot f_{\text{CLK}}$ \\
\hline
{\bf\boldmath $F_{\Delta 2} = 1$}
      & \textttbf{0} & \color{red}{\textttbf{1}} & \textttbf{0} & \textttbf{0} 
      & \textttbf{0} & \color{blue}{\textttbf{1}} & \textttbf{0} & \textttbf{0}
      & $1/4 \cdot f_{\text{CLK}}$ \\
\hline
      & \multicolumn{8}{c|}{\dots} & \\
\hline
{\bf\boldmath $F_{\Delta 7} = 1$}
      & \textttbf{0} & \textttbf{0} & \color{blue}{\textttbf{1}} & \textttbf{0} 
      & \textttbf{0} & \textttbf{0} & \color{red}{\textttbf{1}} & \textttbf{0}
      & $1/4 \cdot f_{\text{CLK}}$ \\
\hline
{\bf\boldmath $F_{\Delta 8} = 1$}
      & \textttbf{0} & \textttbf{0} & \textttbf{0} & \cellcolor{simatablegray}\color{blue}{\textttbf{1}} 
      & \textttbf{0} & \textttbf{0} & \textttbf{0} & \color{red}{\textttbf{1}}
      & $1/4 \cdot f_{\text{CLK}}$ \\
\hline
\end{tabular}
\caption{OR gate with frequency modulation.}
\label{fig:FM_OR_gate_01}
\end{figure}

The standard logic gate and the multiplexor, which are both dark grayed in the figure,
can be implemented in a single LUT, where one input is the synchronization signal,
$\boldsymbol{SYNC}$, and the other two inputs are $\boldsymbol{A}_{\boldsymbol{\Delta}}$
and $\boldsymbol{B}_{\boldsymbol{\Delta}}$.  It can be observed that the LUT output 
will never stay at a constant level; thus, it will evade a UCI analysis.

If an FM logic gate with more that five inputs is to be implemented, then multiple LUTs 
are needed.  In this case it must be ensured that none of the LUT outputs will stay at a
constant level. For example, two LUTs are needed to implement the following 
seven\nobreakdash-input function $\boldsymbol{F}$:
\begin{displaymath}
\boldsymbol{F} = \underbrace{(\boldsymbol{A}_{\boldsymbol{\Delta}} \oplus
\boldsymbol{B}_{\boldsymbol{\Delta}})}_{\displaystyle \text{LUT}_1} +
\underbrace{\boldsymbol{G}(\boldsymbol{X}_{\boldsymbol{\Delta}},
\boldsymbol{Y}_{\boldsymbol{\Delta}}, \boldsymbol{Z}_{\boldsymbol{\Delta}},
\boldsymbol{V}_{\boldsymbol{\Delta}}, \boldsymbol{W}_{\boldsymbol{\Delta}},
\boldsymbol{SYNC})}_{\displaystyle \text{LUT}_2}
\end{displaymath}
If, for some reason, $\boldsymbol{A}_{\boldsymbol{\Delta}} =
\boldsymbol{B}_{\boldsymbol{\Delta}}$, then $\boldsymbol{A}_{\boldsymbol{\Delta}} \oplus
\boldsymbol{B}_{\boldsymbol{\Delta}} \equiv 0$. To avoid a constant value which is
detectable through a UCI analysis, $\boldsymbol{F}$ needs
to be implemented with two FM~gates, which guarantees activity at each FM~gate output.

As mentioned, no activity can be detected through UCI techniques. Only the clock 
should exhibit a very high level of activity. Any other function that exhibits 
such a high level of activity will be suspicious.
In FM~logic, the shortest length of the circular shift register is 4, in which case 
the activity is 25\% for FM~Logic~\textttbf{'0'} and 50\% for FM~Logic~\textttbf{'1'}.
Figure~\ref{fig:FM_logic_gate_01} shows circular shift registers with eight 
stages, which means that the activity is 12.5\% for FM~Logic~\textttbf{'0'} and
25\% for FM~Logic~\textttbf{'1'}. By increasing the length of 
the circular shift registers the activity level decreases serving the purpose of
hiding the Trojan.  An additional benefit of longer circular shift registers is that
they allow for the implementation of multi\nobreakdash-level logic. For example, 
in a circular shift register of length~8, extra information can be 
encoded into Stages~2 and~6.

\begin{figure}[H]
\centering
\includegraphics[scale=0.48]{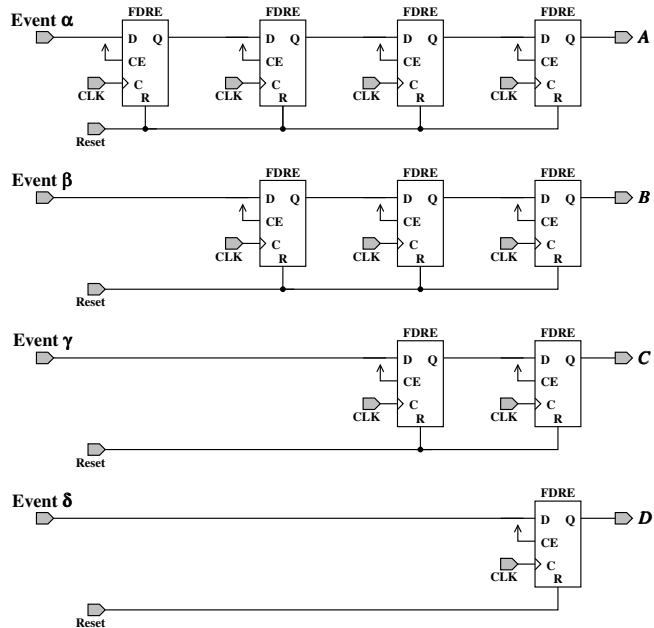}
\caption{Event synchronization circuitry.}
\label{fig:event_synchronization_01}
\end{figure}

Let a string of four consecutive events $\boldsymbol{\alpha}$,
$\boldsymbol{\beta}$, $\boldsymbol{\gamma}$, and $\boldsymbol{\delta}$ be the very 
rare combination that triggers the hardware Trojan.  These events can correspond, for example, 
to four specific operation codes issued in an attacked processor.  
Figure~\ref{fig:event_synchronization_01} presents an event synchronization circuitry, 
which forces the signals $\boldsymbol{A}$, $\boldsymbol{B}$, $\boldsymbol{C}$, and 
$\boldsymbol{D}$ to be simultaneously \textttbf{'1'}
when those four operations are issued in the order 
$\boldsymbol{\alpha}-\boldsymbol{\beta}-\boldsymbol{\gamma}-\boldsymbol{\delta}$.
This activates the function
$\boldsymbol{F}_{\boldsymbol{\Delta}}(\boldsymbol{A}, \boldsymbol{B}, \boldsymbol{C},
\boldsymbol{D})\, =\, \boldsymbol{A}\cdot \boldsymbol{B}\cdot \boldsymbol{C}\cdot 
\boldsymbol{D}$, which in turn triggers the hardware Trojan. 

According to Figure~\ref{fig:FM_logic_gate_01}, it is apparent that
the Trojan's activation is captured into 
$\boldsymbol{F}_{\boldsymbol{\Delta}}(\boldsymbol{A}, \boldsymbol{B}, \boldsymbol{C},
\boldsymbol{D})$ only if signal $\boldsymbol{F}(\boldsymbol{A}, 
\boldsymbol{B}, \boldsymbol{C}, \boldsymbol{D})$ switches to \textttbf{'1'}
synchronously with signal $\boldsymbol{SYNC}$. There are no provisions in 
Figure~\ref{fig:event_synchronization_01} to ensure such a synchronization.
One possible solution is to program the processor ro read the state of the 
$\boldsymbol{SYNC}$ signal and issue the operation codes just in time to achieve the
synchronization.  A second solution is to issue the triggering
string of operation codes a number of times at different time intervals, such that
the synchronization is statistically achieved.

\begin{figure}[H]
\centering
\includegraphics[scale=0.50]{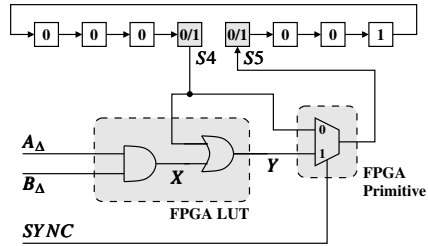}
\caption{Locking AND gate with frequency modulation.}
\label{fig:FM_logic_gate_locking_02}
\end{figure}

In Figure~\ref{fig:FM_logic_gate_01} it is also apparent that the Trojan's activation 
signal, $\boldsymbol{F}_{\boldsymbol{\Delta}}$, remains active for eight clock cycles.
If this activation needs to be locked, then the locking
circuit shown in Figure~\ref{fig:FM_logic_gate_locking_02} can be used.

The proposed logic family supports the implementation of locking gates.  As an 
example, a two\nobreakdash-input locking AND gate with frequency modulation is 
shown in Figure~\ref{fig:FM_logic_gate_locking_02}. In the initial state
$\boldsymbol{S4} = \textttbf{'0'}$.  If $\boldsymbol{A}_{\boldsymbol{\Delta}}$ 
and $\boldsymbol{B}_{\boldsymbol{\Delta}}$ are not simultaneously 
\textttbf{'1'}, then $\boldsymbol{X} = \textttbf{'0'}$,
$\boldsymbol{Y} = \textttbf{'0'}$, and $\boldsymbol{S5} = \textttbf{'0'}$;
therefore, the FM~state \textttbf{'0'} is preserved. If 
$\boldsymbol{A}_{\boldsymbol{\Delta}} = \boldsymbol{B}_{\boldsymbol{\Delta}} = 
\textttbf{'1'}$, then $\boldsymbol{X} = \textttbf{'1'}$, 
$\boldsymbol{Y} = \textttbf{'1'}$, and 
$\boldsymbol{S5} = \textttbf{'1'}$; thus, the FM~state switches to
\textttbf{'1'}.  After seven cycles $\boldsymbol{S4} = \textttbf{'1'}$,
$\boldsymbol{Y} = \textttbf{'1'}$, and $\boldsymbol{S5} = \textttbf{'1'}$,
which means that the FM~state \textttbf{'1'} is preserved independently of
the input signals $\boldsymbol{A}_{\boldsymbol{\Delta}}$ and 
$\boldsymbol{B}_{\boldsymbol{\Delta}}$.

The input signals $\boldsymbol{A}_{\boldsymbol{\Delta}}$ and
$\boldsymbol{B}_{\boldsymbol{\Delta}}$ exhibit FM switching activity: 
$\boldsymbol{S4} = \textttbf{'0'}/\textttbf{'1'}$ (it carries one bit 
of information), $\boldsymbol{S8} = \textttbf{'1'}$, and $\boldsymbol{S1} = 
\boldsymbol{S2} = \boldsymbol{S3} = \boldsymbol{S5} =
\boldsymbol{S6} = \boldsymbol{S7} = \textttbf{'0'}$. As a result, signals 
$\boldsymbol{X}$ and $\boldsymbol{Y}$ are not constant, which allows them 
to escape the UCI detection. Signal $\boldsymbol{SYNC}$ switches periodically,
so it does not pose any detection problem.

\section{Protecting Hardware Trojans with Frequency Modulation to Power 
Consumption Attacks}
\label{sect:protecting_HT_FM_power_attacks}

The deployment of hardware Trojans introduces physical variables.  Their
implementation provides side\nobreakdash-channel information 
(e.g.  power consumption, electromagnetic emissions, propagation delay)
that defenders can use to reveal their existance.  Agrawal \etal~have argued
that this is a serious threat for malicious hardware, since even a hardware 
Trojan that includes only a 16\nobreakdash-bit
counter, an 8\nobreakdash-bit sequential comparator, and a 3\nobreakdash-bit
combinational comparator is large enough to consume the power necessary 
for its detection\cite{Agrawal_May07}.  These findings are in line with 
Potkonjak \etal, who have shown that gate\nobreakdash-level characterization 
based on timing and power consumption measurements can help with the detection of 
hardware Trojans\cite{Potkonjak_Jul09}.
It has also been reported that a network of Ring Oscillators, which are similar
to those used in the proposed logic family with frequency modulation, 
can provide sufficient sensitivity to detect changes in power consumption generated 
by malicious circuitry\cite{Zhang_Mar11}.  In addition,
the field\nobreakdash-programmable gate arrays are notable for their
large power consumption.  As a consequence, reconfigurable implementations 
are even more vulnerable than ASICs to side\nobreakdash-channel 
analysis\cite{Ors_Sep03,Standaert_AugSep04}.

Given the above\nobreakdash-mentioned considerations it is apparent that
building a stealthy hardware Trojan requires the elimination of the relationships
between data and power consumption, and between operations and power consumption.
There are two ways to achieve this\cite{Mangard_10}:
(i)~\textbf{hiding} (also called \textbf{concealing}), which makes the 
power consumption independent relative to the processed data and/or operations, 
and (ii)~\textbf{masking}, which randomizes the power consumption.  In this paper, 
only the hiding technique is used in securing the hardware Trojan.

Recall that the CMOS power consumption has two components: 
(i)~\textbf{dynamic} and (ii)~\textbf{static} (also referred to as \textbf{leakage}).
In order for a hardware Trojan to be stealthy, it must be robust enough to evade
analyses based on each and every power consumption component. A common approach
is to use Dual\nobreakdash-Rail Logic (DRL)\cite{Mangard_10}, which balances the dynamic power
consumption into a constant value through differential encoding in which the 
information is encoded with direct and complementary signals.  DRL operates 
in two alternating phases: \textsl{precharge}, during which both the direct 
and complementary signals are set to a common Low value, and \textsl{evaluation}, 
during which either the direct signal (which encodes logic '1') or the complementary 
signal (which encodes logic '0') will perform a transition to a High value. This way, 
exactly one High\nobreakdash-to\nobreakdash-Low transition during the precharge phase
and one Low\nobreakdash-to\nobreakdash-High transition during the evaluation phase
occur in this two\nobreakdash-phase operation irrespective of the logic state 
('0' or '1') being encoded. The consequence is a constant
dynamic power consumption.

Measures to increase robustness against analyses based on static power consumption
have also been proposed\cite{Halak_Jul15}. The main idea in concealing the leakage is to
ensure that the number of states in logic '0' is always equal to the number of states in
logic '1'. Since the circuit exhibits symmetry, the leakage power consumption is constant.

\begin{figure}[H]
\centering
\includegraphics[scale=0.62]{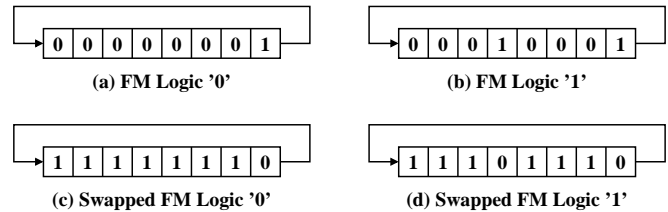}
\caption{Power consumption concealment.}
\label{fig:power_concealment_01}
\end{figure}

The power consumption concealment is achieved through hardware replication.  
Figure~\ref{fig:power_concealment_01} shows that the circular shift register is
replicated such that for each CSR storing FM Logic '0' (a) there is a dual CSR storing FM Logic
'1' (b).  The CSRs (a) and (b) are replicated into (c) and (d) such that the
number of states in standard logic '0' equals the number of states in standard logic '1'.
This conceals the leakage power consumption.  By providing the dual FM logic, the
number of standard transitions 0\nobreakdash-to\nobreakdash-1 is always 6, and the
number of standard transitions 1\nobreakdash-to\nobreakdash-0 is always 6. This conceals
the dynamic power consumption.  Also, both frequencies ($1/4\cdot f_{\text{CLK}}$ and
$1/8\cdot f_{\text{CLK}}$) will always be present in the side\nobreakdash-channel
spectrum, which will defuse attacks based on spectrum analysis.

The proposed implementation is significantly simpler than the
dual\nobreakdash-rail logic since it lacks the precharge and the evaluation phases. This is
possible since the FM~logic is based on circular shift registers,
which exhibit repetitive switching behavior. Since the standard logic gate and the 
multiplexor, which are both dark grayed in Figure~\ref{fig:FM_logic_gate_01}, can be 
implemented in a single LUT, their power consumption can be easily concealed by
implementing the dual function in the dual LUT.

After the hardware Trojan is activated, the concealment of the power consumption is
no longer needed. The replicas used in protecting the Trojan can now be used to implement
the payload. In a first scenario, replicas \ref{fig:power_concealment_01}\nobreakdash-(b),
\ref{fig:power_concealment_01}\nobreakdash-(c), and
\ref{fig:power_concealment_01}\nobreakdash-(d) can be disabled upon the activation of the
Trojan. This will make power consumption dependent on the data and operations
implemented with frequency\nobreakdash-modulated logic, which in turn will allow the 
attacker to retrieve secret information through this side channel.  In a second
scenario, replicas \ref{fig:power_concealment_01}\nobreakdash-(c) and
\ref{fig:power_concealment_01}\nobreakdash-(d) are disabled, but both replicas
\ref{fig:power_concealment_01}\nobreakdash-(a) and
\ref{fig:power_concealment_01}\nobreakdash-(b) hold the same FM~logic value.
This will double the dependence of the power consumption on data and operations, with
beneficial effects in implementing the payload.

\section{Detecting Hardware Trojans Built with Frequency-Modulated Logic}
\label{sect:detecting_HT_FM}

It has been shown that the hardware Trojans built with frequency\nobreakdash-modulated
logic are robust to tests based on unused\nobreakdash-circuit identification and
side\nobreakdash-channel analysis. A natural question at this point is how such Trojans
can be detected and/or neutralized given their new behavior.

The logic with Frequency Modulation (FM) require a large number of circular shift 
registers, where a standard logic gate (or a LUT), a 2:1~multiplexer, and a circular
shift register are needed for each FM~gate.  This pattern, which can be regarded as a
signature of this type of logic, can be the subject of detection attempted at the RTL 
level. However, if the HDL code is encrypted, the detection can only be based
on side\nobreakdash-channel analyses.

It should be observed that the robustness of dual\nobreakdash-rail logic circuits 
mapped onto FPGAs might not be very good, since the routing imbalance in reconfigurable
arrays lowers the effectiveness of this type of logic\cite{Tiri_Feb04}.  This observation
suggests that the hardware Trojans with frequency modulation should be implemented
with FPGA primitives (\textsl{LUT6}, \textsl{FDRE}, \textsl{FDSE},
\textsl{MUXF7}, and \textsl{MUXF8}) rather than behavioral HDL code, and the placement 
should be controlled with location attributes and constraints. If the resulting Trojan is
robust to side\nobreakdash-channel analysis, it is legitimate to investigate other
opportunities in defending the circuit.

If the Trojan's triggering circuitry cannot be detected, then the only chance to defend
the circuit is by neutralizing the Trojan's payload.  This can be achieved, for example,
through a Counter\nobreakdash-Trojan deployed on the FPGA by the defender.
The Counter\nobreakdash-Trojan will perform a spectral analysis to
determine if oscillations with frequencies that are large fractions of $f_{\text{CLK}}$ 
exist. Then it will trigger its own oscillations on the
same frequencies to jam the original payload, so that the attacker is no longer able to
communicate with the Trojan. Other proposed techniques are also
worth investigating\cite{Zhou_May16}.

\section{Conclusion}
\label{sect:conclusion}

Defense measures at the RTL,
side\nobreakdash-channel analysis, and Counter\nobreakdash-Trojan levels have been
described to mitigate the risk of a novel security vulnerability disclosure.






\end{document}

%% file: import_packages_and_definitions.tex
\usepackage{times}
\usepackage[T1]{fontenc}
\usepackage{textcomp}
\usepackage{pslatex}
\usepackage{amsmath}
\usepackage{amsfonts}
\usepackage{amssymb}
\usepackage{amsthm}
\usepackage{mathtools}
\usepackage{commath}
\usepackage{mathdots}
\usepackage{algorithm}
\usepackage{algorithmic}
\usepackage{graphicx}
\usepackage{subfig}
\usepackage{wrapfig}
\usepackage{url}
\usepackage{lscape}
\usepackage{cite}
\usepackage{color}
\usepackage{colortbl}
\usepackage{bm}
\usepackage{multirow}
\usepackage{float}
\usepackage{rotating}
\usepackage{array}
\usepackage{fixltx2e}		
\usepackage{listings}
\usepackage[mathscr]{euscript}
\usepackage{fancyvrb} 
\usepackage{eurosym}
\usepackage{wasysym}  
\usepackage{gensymb}  
\usepackage{comment}
\usepackage[none]{hyphenat}
\usepackage[makeroom]{cancel}
\usepackage{datetime}
\usdate
\usepackage{ulem}
\usepackage{soul}
\usepackage{logpap}
\usepackage{slashbox}  
\usepackage{calc}
\usepackage{fancyhdr,lastpage}
\usepackage{lipsum}
\usepackage{cuted}
\usepackage[shortlabels]{enumitem}
\setlistdepth{9}
\newlist{sima_itemize}{itemize}{9}
\setlist[sima_itemize]{labelsep=7pt,leftmargin=23pt}
\setlist[sima_itemize,1]{label=$\bullet$,leftmargin=11pt}
\setlist[sima_itemize,2]{label=$\ast$}
\setlist[sima_itemize,3]{label=$\circ$}
\setlist[sima_itemize,4]{label=$\star$}
\setlist[sima_itemize,5]{label=$\diamond$}
\setlist[sima_itemize,6]{label=$\triangleright$}
\setlist[sima_itemize,7]{label=$\cdot$}
\setlist[sima_itemize,8]{label=$\cdot$}
\setlist[sima_itemize,9]{label=$\cdot$}
\usepackage{cleveref}
\usepackage{arydshln}
\definecolor{simagray}{gray}{0.95}
\definecolor{simared}{rgb}{1.0,0,0}
\definecolor{simagreen}{rgb}{0.0,0.6,0.0}
\definecolor{simatablegray}{gray}{0.80}
\newlength{\epswidth}
\newcommand{\etal}{\textsl{et al.}}

\newcommand{\textttbf}[1]{\textbf{\texttt{#1}}}

\newcommand\longvisiblespace[1][.3em]{%
  \mbox{\kern.06em\vrule height.3ex}%
  \vbox{\hrule width#1}%
  \hbox{\vrule height.3ex}}

\crefname{equation}{Equation}{Equations}
\crefname{table}{Table}{Tables}